\documentclass[a4paper]{spie}  %>>> use this instead for A4 paper
\usepackage{amsmath,amsfonts,amssymb}
\usepackage{graphicx}
\usepackage[colorlinks=true, allcolors=blue]{hyperref}

\title{Commissioning the AIP's ultrafast laser Inscription facility: defining the parameter space for type-I waveguide fabrication}

\author[a,b]{Rima Islam$^*$}
\author[c]{Aline N. Dinkelaker}
\author[a]{Abani Shankar Nayak}
\author[a]{Kalaga Madhav}
\author[a]{Martin Roth}

\affil[a]{Leibniz Institute for Astrophysics Potsdam (AIP), Potsdam, Germany}
\affil[b]{Friedrich-Schiller-Universität Jena, Jena, Germany}
\affil[c]{Corning Optical Communications GmbH $\& $ Co. KG, Berlin, Germany}

\authorinfo{%Further author information: (Send correspondence to Rima Islam)\\
$^*$Rima Islam: E-mail: rima.islam@uni-jena.de}

\usepackage[a4paper, left=19.3mm, right=19.3mm, top=25mm, bottom =25mm]{geometry}

\begin{document} 
\maketitle

%----------------------

%----------------------

\begin{abstract}
Astrophotonics offers a compact, stable alternative to bulk optics in traditional astronomical instrumentation, spanning high-precision spectroscopy to high-contrast interferometry. However, the strict requirements for throughput, accuracy, precision, and stability require tailored manufacturing processes that go beyond standard telecommunications technology. This work presents the iterative experimental and statistical methods used to determine the optimal laser-writing conditions for passive waveguides in silica glass using a femtosecond laser inscription setup. We systematically explored the parameter space by varying pulse energy, repetition rate, translation speed, and scan numbers to identify the optimal conditions for Type-1 waveguides, characterized by a positive refractive-index modification in their cores. 

\noindent We report on the refinement of these parameters to manufacture single-mode, high-throughput waveguides that operate in the astronomical J-band (1300-1400 nm). A key focus of this optimization is to match the mode field diameter (MFD) to the SMF28 fibre and minimize insertion losses. This development is critical for astrophotonics devices, such as photonic reformatters, beam combiners, and pupil remappers, as well as post-fabrication correction of phase errors in photonic components fabricated using traditional photolithographic methods. Some of these components will be used for future on-sky validation at the Calar Alto Observatory and the CHARA array.

% --------------
% This document is prepared using LaTeX2e\cite{Lamport94} and shows the desired format and appearance of a manuscript prepared for the Proceedings of the SPIE.\footnote{The basic format was developed in 1995 by Rick Hermann (SPIE) and Ken Hanson (Los Alamos National Lab.).} It contains general formatting instructions and hints about how to use LaTeX.  The LaTeX source file that produced this document, {\ttfamily article.tex} (Version 3.4), provides a template, used in conjunction with {\ttfamily spie.cls} (Version 3.4). These files are available on the Internet at \url{https://www.overleaf.com}.  The font used throughout is the LaTeX default font, Computer Modern Roman, which is equivalent to the Times Roman font available on many systems.
% --------------

\end{abstract}


%---------------------
% future work 
% needs to verify the refractive index in the future and also, check the circularity 
%---------------------

\section{INTRODUCTION}
\label{sec:intro}  % \label{} allows reference to this section

\textcolor{black}{Astrophotonics has emerged as a transformative technology for modern observational astronomy, enabling the miniaturization of complex optical instruments into compact, stable, and highly integrated photonic platforms \cite{Dinkelaker2024AstrophotonicTechnologies}. By leveraging technologies originally developed for the telecommunications industry, astrophotonics offers significant advantages in terms of optical stability, scalability, reduced alignment complexity, and enhanced performance, addressing many of the challenges faced by next-generation astronomical facilities. As astronomical observatories continue to pursue higher angular resolution, greater sensitivity, and improved measurement precision, photonic technologies are becoming increasingly important for meeting these demanding instrumental requirements \cite{norris2024astrophotonics}.} 

\noindent \textcolor{black}{Over the past decade, astrophotonic devices have matured from laboratory demonstrations to operational components within astronomical instruments. Key examples include photonic lanterns for efficient multimode-to-single-mode conversion \cite{PL2015}, fiber Bragg gratings for atmospheric OH-line suppression \cite{ULI-OH-2018}, astro-frequency combs for ultra-precise wavelength calibration, integrated photonic spectrographs for compact spectroscopy \cite{roth2023}, and beam combiners for long-baseline interferometry \cite{sanny+nott-iii+2026, sanny+spie+2026}. These devices provide unique capabilities that are difficult or impossible to achieve using conventional bulk-optics approaches while simultaneously reducing instrument size, weight, and environmental sensitivity. }

\noindent \textcolor{black}{Several fabrication technologies have been explored for astrophotonic device development, including planar lithographic processes and direct laser-writing techniques \cite{Dinkelaker2024AstrophotonicTechnologies}. Among these, Ultrafast Laser Inscription (ULI) has attracted considerable interest for its ability to directly write three-dimensional waveguide structures inside transparent dielectric materials without requiring cleanroom facilities \cite{gross2015}. The technique uses tightly focused femtosecond laser pulses to induce localized refractive index modifications, enabling rapid prototyping and fabrication of complex photonic circuits at relatively low infrastructure costs. Furthermore, the intrinsic three-dimensional design freedom of ULI enables sophisticated routing geometries, waveguide crossovers, and beam-combining architectures that are challenging to achieve with conventional planar beam combiners produced by lithography. These advantages have established ULI as a leading technology for astronomical integrated optics, particularly within the interferometry community \cite{sanny+2T-nuller+2026,sanny+nott-iii+2026,Harris2018NAIR}.}

\noindent \textcolor{black}{At the Leibniz Institute for Astrophysics Potsdam (AIP), astrophotonic technologies are being actively developed to support future high-angular-resolution astronomical instrumentation. Recent achievements include the development of K-band integrated-optics beam combiners for the CHARA array through the CHARIOT project, demonstrating the scientific potential of photonic interferometric instrumentation \cite{Mayer2024CHARIOT}. As astrophotonic activities continue to expand within AIP, establishing an in-house ULI fabrication capability has become strategically important to accelerate device development cycles, enable rapid prototyping, and reduce dependence on external fabrication facilities.}

% \noindent \textcolor{black}{At the Leibniz Institute for Astrophysics Potsdam (AIP), astrophotonic technologies are being actively developed to support future high-angular-resolution astronomical instrumentation. Recent achievements include the development of H-band integrated-optics beam combiners for the CHARA array through the CHARIOT project, demonstrating the scientific potential of photonic interferometric instrumentation \cite{Mayer2024CHARIOT}. As astrophotonic activities continue to expand within AIP, establishing an in-house ULI fabrication capability has become strategically important to accelerate device development cycles, enable rapid prototyping, and reduce dependence on external fabrication facilities.}

\noindent \textcolor{black}{The present work, therefore, focuses on developing the first in-house ULI fabrication capability at AIP using an available 520 nm femtosecond laser platform. The initial objective is to establish and optimize laser inscription parameters for fabricating low-loss waveguides operating in the near-infrared, with particular emphasis on the J-band (1300--1400 nm). Efficient coupling to standard single-mode fibers, such as SMF-28, is a critical requirement, as these fibers are widely used for fan-in and fan-out to astrophotonic devices and for interfacing with existing J-band photonic components under development at AIP\cite{Dinkelaker2021SixTelescope}. The optimization process investigates the effects of laser power, pulse overlap, scan speed, and writing geometry on the resulting waveguide morphology and optical performance. Beyond the fabrication of individual waveguides, this effort serves as the foundation for a broader astrophotonics development program at AIP. The long-term objective is to realize complete passive integrated-optics beam combiners fabricated entirely in-house using ULI technology. Such devices would support future astronomical interferometry instruments and pave the way for on-sky demonstrations and commissioning campaigns, following the successful examples of the GRAVITY \cite {gravity2018} and Asgard/NOTT \cite {sanny+nott-iii+2026} beam-combining instruments.}

\section{Experimental Setups}
\label{sec:setup}  % \label{} allows reference to this section

\subsection{ULI facility}
\begin{figure}[ht!]
    \centering
    \includegraphics[width=0.75\linewidth]{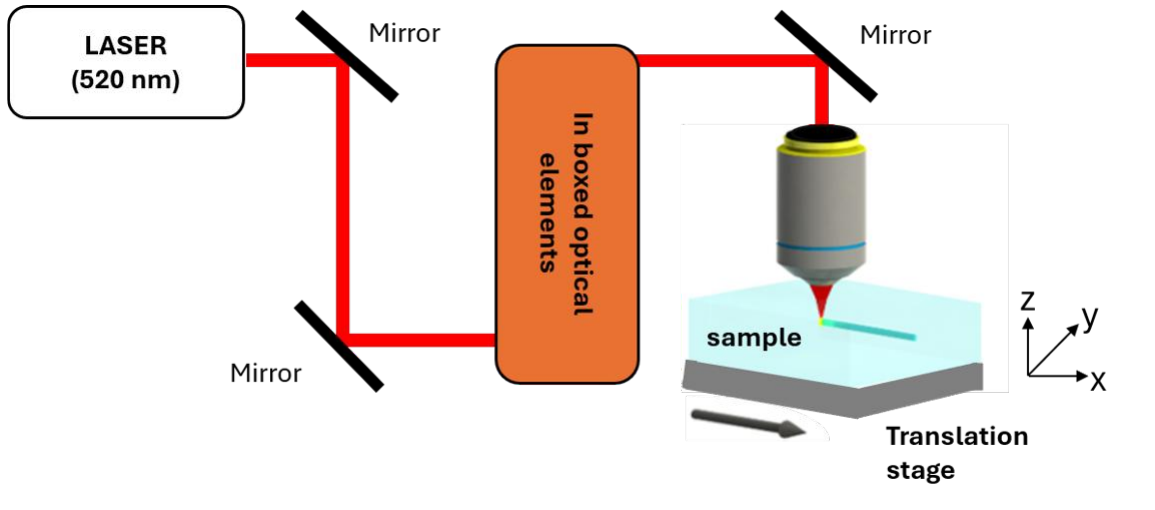}
    \caption{
    \textcolor{black}{Schematic representation of the ultrafast laser inscription (ULI) system used for waveguide fabrication, with corresponding components of the physical setup. The system features a 520 nm femtosecond laser source (Spirit One SPOne-8-SHG, top left) coupled to an optical assembly (orange box) containing power-control elements. The laser beam is focused into the substrate through a microscope objective, while motorized translation stages (bottom right) provide automated three-dimensional sample positioning for waveguide inscription.}
    }
    \label{fig:ULI-system}
\end{figure}

\textcolor{black}{As a first step toward establishing an in-house ultrafast laser inscription (ULI) fabrication capability at AIP, waveguides were fabricated using a commercial FemtoFBG ULI system (MKS Instruments). The system comprises a femtosecond laser source (Spirit One, SPOne-8-SHG) and high-precision motorized translation stages (Newport XMS100 for the $x$- and $y$-axes, and Newport VP-5ZA for the $z$-axis), enabling automated sample positioning and waveguide inscription. A schematic representation of the setup is shown in Figure~\ref{fig:ULI-system}. Waveguide layouts and stage control were implemented through the $\mu$FAB software package, which allows the design and automated fabrication of photonic structures.} 

% \textcolor{black}{As a first step toward establishing an in-house ultrafast laser inscription (ULI) fabrication capability at AIP, waveguides were fabricated using a commercial FemtoFBG ULI system (MKS Instruments). The system comprises a femtosecond laser source (Spirit One, SPOne-8-SHG) and high-precision motorized translation stages (Newport XMS100 for the $x$- and $y$-axes, and Newport VP-5ZA for the $z$-axis), enabling automated three-dimensional sample positioning and waveguide inscription. A schematic representation of the setup is shown in Figure~\ref{fig:ULI-system}. Waveguide layouts and stage control were implemented through the $\mu$FAB software package, which allows the design and automated fabrication of photonic structures.} 

\noindent \textcolor{black}{The laser source can operate either at its fundamental wavelength of 1040 nm or at the second-harmonic wavelength of 520 nm. In the present study, all fabrication experiments were performed using the 520 nm output. The objective of this initial investigation was to identify the optimal inscription conditions for producing low-loss, single-mode waveguides suitable for operation in the astronomical J-band.}

% \textcolor{black}{The laser source can operate either at its fundamental wavelength of $1040,\mathrm{nm}$ or at the second-harmonic wavelength of $520,\mathrm{nm}$. In the present study, all fabrication experiments were performed using the $520,\mathrm{nm}$ output. The objective of this initial investigation was to identify the optimal inscription conditions for producing low-loss, single-mode waveguides suitable for operation in the astronomical J-band.}

\noindent \textcolor{black}{To systematically optimize the fabrication process, a range of laser-writing parameters was investigated. The primary inscription parameters include the pulse energy, $E_p$ [nJ], repetition rate, $R_p$ [kHz], average laser power, $P_{\mathrm{avg}}$ [mW], laser wavelength, $\lambda$ [nm], writing-objective numerical aperture, $W_{\mathrm{obj}}$ [NA], inscription depth, $z$ [$\mu$m], and focal spot diameter, $D_s$ [$\mu$m]. In addition, the interaction between the laser and material during inscription is governed by process-dependent parameters such as the scan speed, $V_s$ [mm/s], and the number of pulses deposited per unit length, $N_p$. These quantities can be combined into the representative net fluence (RNF) [$\mu$J/$\mu$m$^2$], which provides a convenient metric for comparing the accumulated energy delivered to the material under different writing conditions.} 

% \textcolor{black}{To systematically optimize the fabrication process, a range of laser-writing parameters was investigated. The primary inscription parameters include the pulse energy, $E_p$ [nJ], repetition rate, $R_p$ [kHz], average laser power, $P_{\mathrm{avg}}$ [mW], laser wavelength, $\lambda$ [nm], writing-objective numerical aperture, $W_{\mathrm{obj}}$ [NA], inscription depth, $z$ [$\mu$m], and focal spot diameter, $D_s$ [$\mu$m]. In addition, the interaction between the laser and material during inscription is governed by process-dependent parameters such as the scan speed, $V_s$ [mm/s], and the number of pulses deposited per unit length, $N_p$. These quantities can be combined into the representative net fluence (RNF) [$\mu$J/$\mu$m$^2$], which provides a convenient metric for comparing the accumulated energy delivered to the material under different writing conditions. Consequently, the optimization strategy adopted in this work focused on exploring the influence of these parameters on the resulting waveguide morphology and optical performance in order to identify the conditions that support stable single-mode guidance.
% } 

\subsection{Phase Contrast Microscope}
\textcolor{black}{Following fabrication with the ULI system, straight waveguide structures were examined using phase-contrast microscopy (PCM) as a rapid, non-destructive method \cite{Murphy2012, Ockenga2023}. PCM converts optical phase variations arising from local changes in refractive index and density into intensity contrast.} 

% \textcolor{black}{Following fabrication with the ULI system, straight waveguide structures were examined using phase-contrast microscopy (PCM) as a rapid, non-destructive method for assessing laser-induced modifications within the substrate \cite{Murphy2012, Ockenga2023}. PCM converts optical phase variations arising from local changes in refractive index and density into intensity contrast, enabling visualization of the inscribed structures prior to optical characterization.} 

\begin{figure}[ht!]
    \centering
    \includegraphics[width=0.85\linewidth]{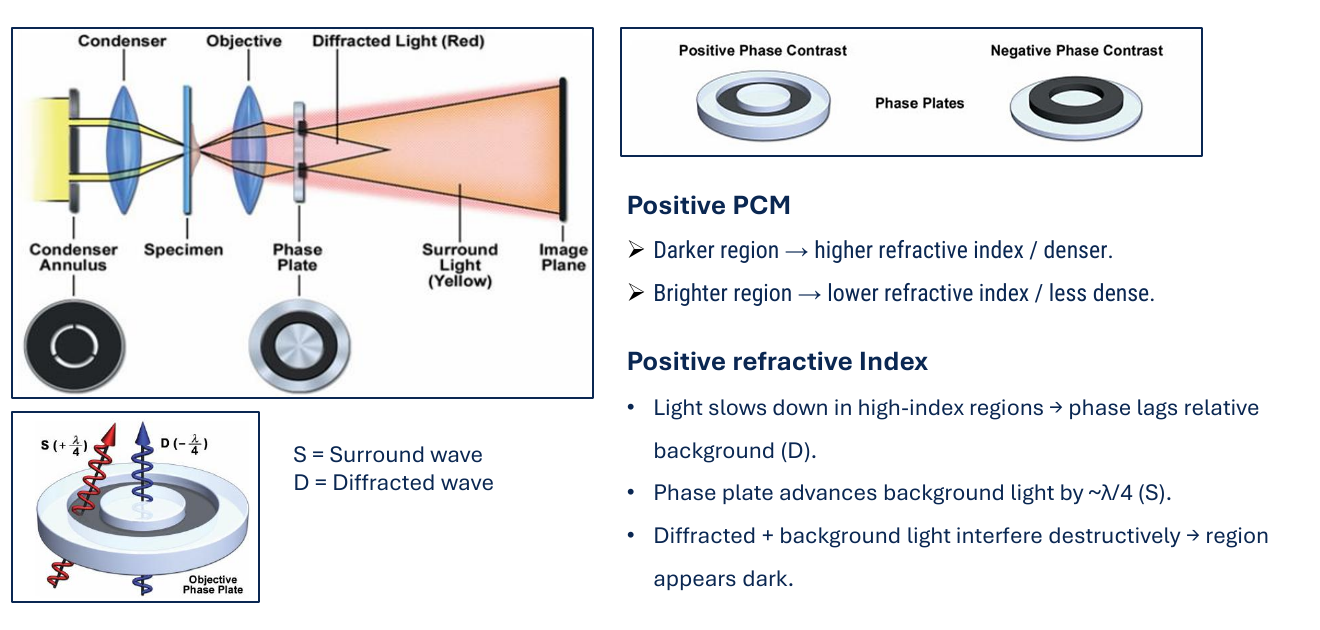}
    \caption{
    \textcolor{black}{Principle of positive phase-contrast microscopy (PCM) used for the preliminary inspection of ultrafast laser-inscribed waveguides. The technique converts phase variations caused by local refractive-index modifications within a transparent material into measurable intensity differences through the use of a condenser annulus and phase plate. In positive phase-contrast imaging, regions with a higher refractive index (or increased material density) induce a larger phase delay and appear darker than the surrounding background, whereas regions with a lower refractive index (or reduced density) appear brighter (image adapted from \cite{Murphy2012, Ockenga2023}).}
    }
    \label{fig:PCM-system}
\end{figure}

% {Principle of positive phase-contrast microscopy (PCM) used for the preliminary inspection of ultrafast laser-inscribed waveguides. The technique converts phase variations caused by local refractive-index modifications within a transparent material into measurable intensity differences through the use of a condenser annulus and phase plate. In positive phase-contrast imaging, regions with a higher refractive index (or increased material density) induce a larger phase delay and appear darker than the surrounding background, whereas regions with a lower refractive index (or reduced density) appear brighter. Although PCM does not provide a direct quantitative measurement of refractive-index change, it offers a rapid and non-destructive method for 

\noindent \textcolor{black}{As illustrated in Figure~\ref{fig:PCM-system}, positive phase-contrast imaging was used throughout this work. Under this convention, regions appearing darker than the surrounding background are interpreted as positive refractive-index modifications associated with localized material densification, whereas brighter regions are associated with negative refractive-index modifications resulting from reduced material density. Representative material responses observed during the inscription parameter study are summarized in Figure~\ref{fig:materials response summary}.} 

\noindent \textcolor{black}{Although PCM does not provide a direct quantitative measurement of refractive-index change, it offers a valuable first-order assessment of the inscription quality and uniformity of the fabricated waveguides. Consequently, PCM served as an intermediate screening step in the optimization workflow (Figure~\ref{fig:workflow}), enabling the rapid identification of unsuitable writing conditions before.}

% enabling the rapid identification of unsuitable writing conditions before proceeding to the more time-intensive near-infrared optical characterization required to evaluate guiding performance

% \textcolor{black}{Although PCM does not provide a direct quantitative measurement of refractive-index change, it offers a valuable first-order assessment of the inscription quality, morphology, and uniformity of the fabricated waveguides. Consequently, PCM served as an intermediate screening step in the optimization workflow (Figure~\ref{fig:workflow}), enabling the rapid identification of unsuitable writing conditions before proceeding to the more time-intensive near-infrared optical characterization required to evaluate guiding performance and single-mode operation.} 

%to identify they are  Following the fabrication of the structures or waveguides, a positive phase-contrast microscope (PCM) was initially used to qualitatively inspect the laser-induced modifications and assess the nature of the refractive index changes within the material. In positive-phase-contrast microscopy, regions appearing darker than the background are interpreted as having a higher density and are referred to in this report as positive refractive index regions, whereas regions appearing brighter than the background are interpreted as lower density and are referred to as negative refractive index regions (\cite{Murphy2012,Ockenga2023}).

\subsection{Near-IR bench}
\textcolor{black}{Following the preliminary inspection using a microscope, the optical performance of the fabricated straight waveguides was evaluated using a near-infrared (NIR) characterization bench to identify waveguides supporting single-mode. The fabrication parameters corresponding to waveguides exhibiting stable single-mode behavior were subsequently considered as the experimentally optimized ULI parameters for the investigated material.}

\begin{figure}[ht!]
    \centering
    \includegraphics[width=0.70\linewidth]{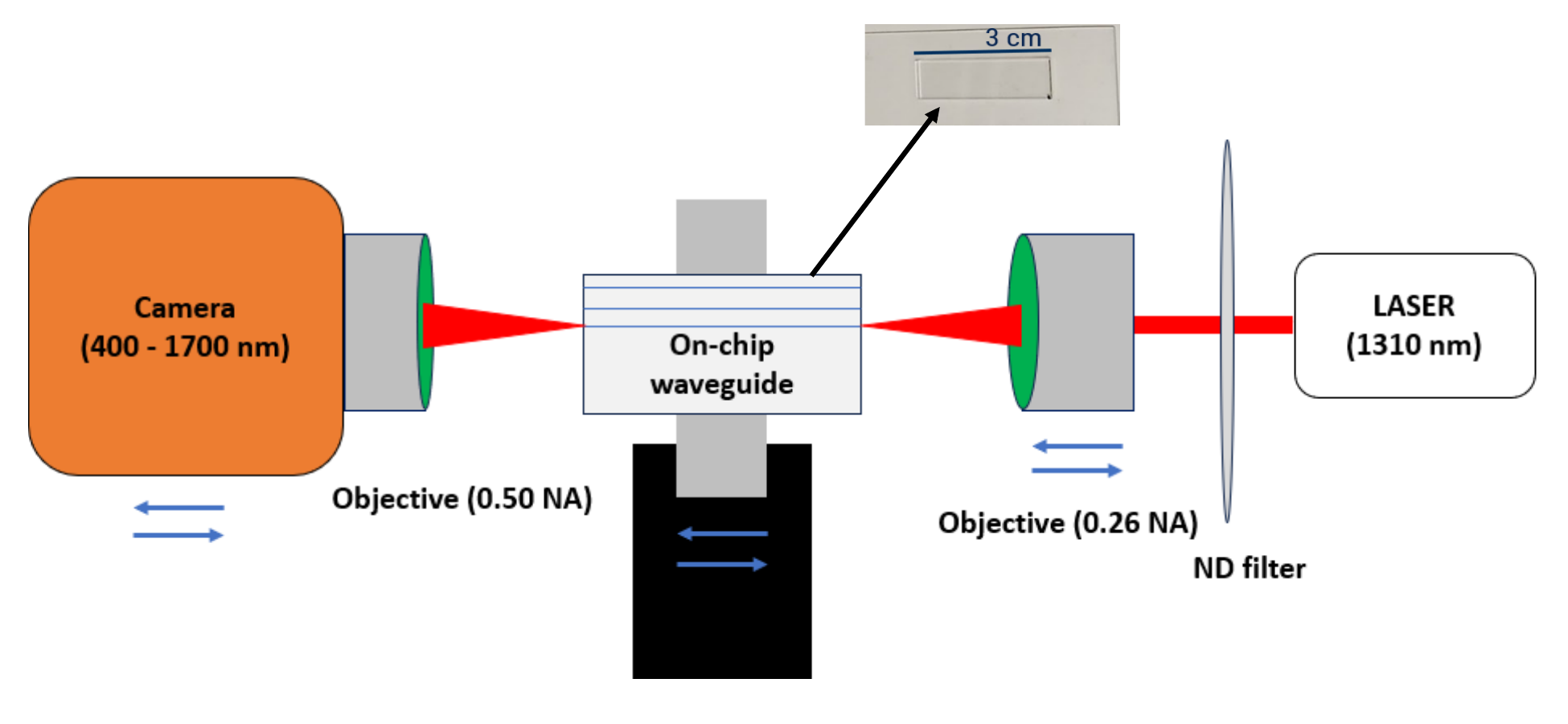}
    \caption{
    \textcolor{black}{Near-infrared characterization setup for evaluating ULI-fabricated waveguides. Light from a 1310 nm laser source is coupled into the waveguide using an objective lens (NA = 0.26), and the output mode is imaged through a collection objective (NA = 0.50) onto an infrared camera. Precision translation stages enable alignment of the coupling optics and waveguide chip. The inset shows a representative 3 cm straight waveguide sample.}
    }
    \label{fig:IR-bench}
\end{figure}

\noindent \textcolor{black}{The characterization setup is shown in Figure~\ref{fig:IR-bench}. An infrared laser source centered at 1310 nm, within the astronomical J-band, was coupled into the waveguide through an objective lens with focal length $f = 20$ mm, working distance of 30.5 mm, and numerical aperture, NA = 0.26. The output light from the waveguide was collected and collimated using a second objective lens with focal length $f = 2$ mm, working distance 17 mm, and NA = 0.50. The resulting near-field mode profile was recorded using an infrared camera.}

\noindent \textcolor{black}{To verify single-mode operation, each straight waveguide was precisely aligned using the translational stage of the chip holder, which enabled controlled horizontal and vertical adjustments of the coupling positions. The mode field profile was observed as the launch conditions were varied using the translation stage, a typical procedure for verifying the modal behavior of the waveguides \cite{sanny2024}. Waveguides that consistently maintained a single, stable fundamental mode under these perturbations were classified as single-mode and used to identify the optimal inscription parameters.} 

% Waveguides that consistently maintained a single, stable fundamental mode under these perturbations were classified as single-mode and used to identify the optimal inscription parameters.

%. A $10\times$ input objective with a numerical aperture (NA) of 0.26 was employed for light injection, while the output light was collected using a $100\times$ objective with an NA of 0.50. The output mode profiles were recorded using an infrared camera sensitive over the 0.4--1.7\,\,$\mu$m wavelength range. The camera, chip stage, and input objective were mounted on translation stages to enable precise alignment during coupling and characterization. Neutral density (ND) filters were inserted as needed to attenuate the optical power and prevent camera saturation.

%-------------
% \textbf{Phase contrast micrsocopy}

% For preliminary characterization to inspect the type of modification and the ULI-written structures, a phase-contrast microscope was used.  Although it does not provide a complete or fully quantitative assessment of ULI-fabricated structures, since image contrast can be affected by factors such as sample thickness, and scattering. However, it offers several practical advantages. In particular, PCM enables rapid, qualitative evaluation, which is especially valuable during the initial exploration of different parameter combinations. Furthermore, the technique allows detection of subtle refractive index changes through induced phase shifts, without the need for staining. In this report, Type~I modification refers to a positive refractive index change, whereas Type~II modification refers to a negative refractive index change.
%-------------
% \section{Mathematical equation}
% \label{sec: Mathematical equation}
% Astrophotonics is emerging as a key technology for modern observational astronomy by replacing bulky optical systems with compact, efficient, and highly integrated photonic devices (Dinkelaker 2024; Jovanovic 2023). Originally derived from 

\section{Results and Discussion}
\label{sec:Methodology and procedure}

\begin{figure} [ht!]
    \centering
    \includegraphics[width=0.7\linewidth]{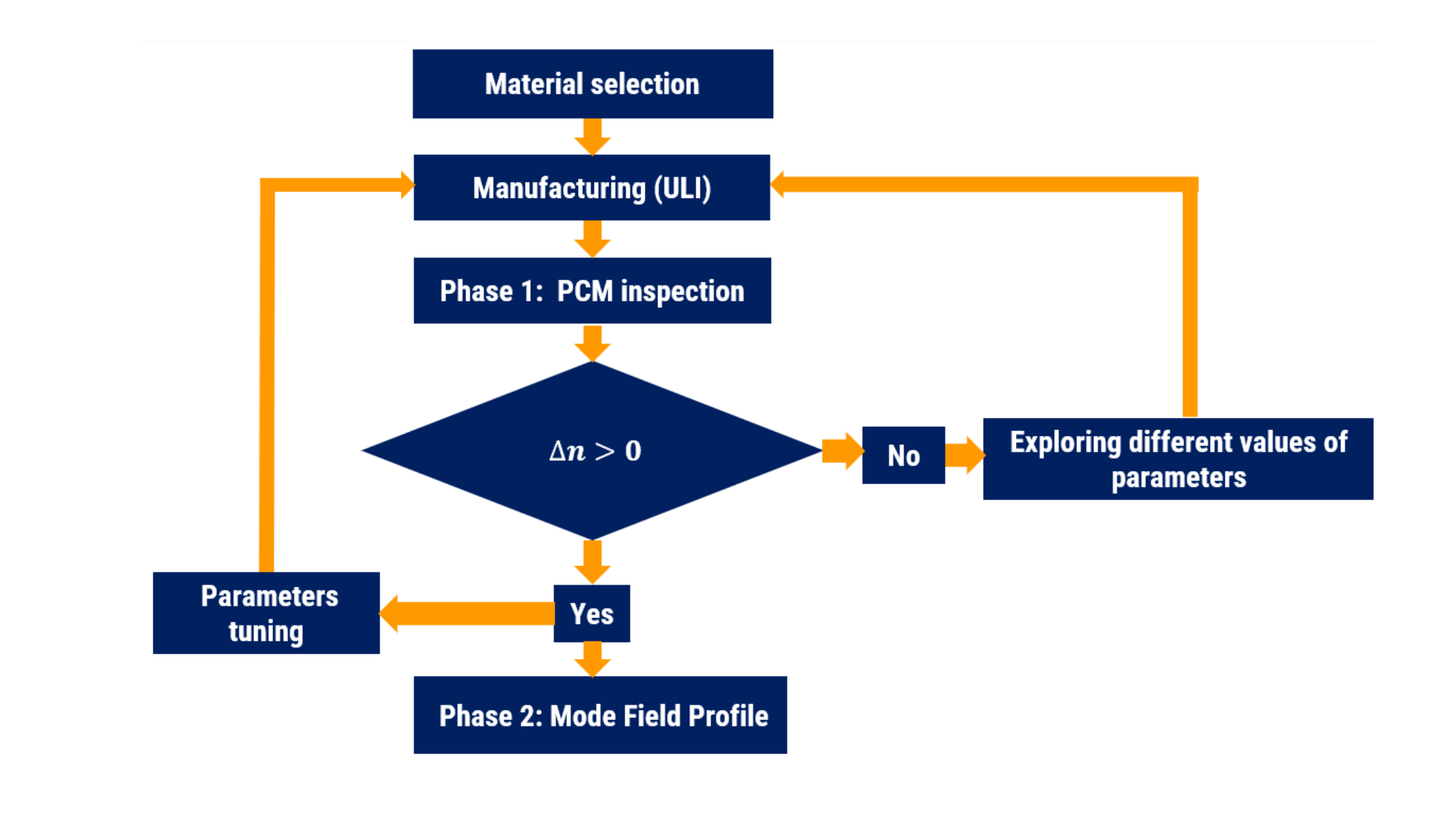}
    \caption{Experimental workflow illustrating material selection, ULI fabrication, ULI parameter optimization, and the successive characterization stages employed for waveguide development.}
    \label{fig:workflow}
\end{figure}

\textcolor{black}{The experimental study was organized into sequential stages of fabrication, inspection, and optical characterization, as summarized in the workflow shown in Figure~\ref{fig:workflow}. Following ULI fabrication, three candidate substrate materials, Borofloat 33, Fused silica, and Eagle XG, were systematically evaluated to determine their suitability for near-infrared waveguide fabrication for this work.}

% \textcolor{black}{The experimental campaign was organized into sequential stages of fabrication, inspection, and optical characterization, as summarized in the workflow shown in Figure~\ref{fig:workflow}. Following ULI fabrication, three candidate substrate materials—Borofloat 33, fused silica, and Eagle XG—were systematically evaluated to determine their suitability for near-infrared waveguide fabrication.}

\begin{figure} [ht!]
    \centering
    \includegraphics[width=0.7\linewidth]{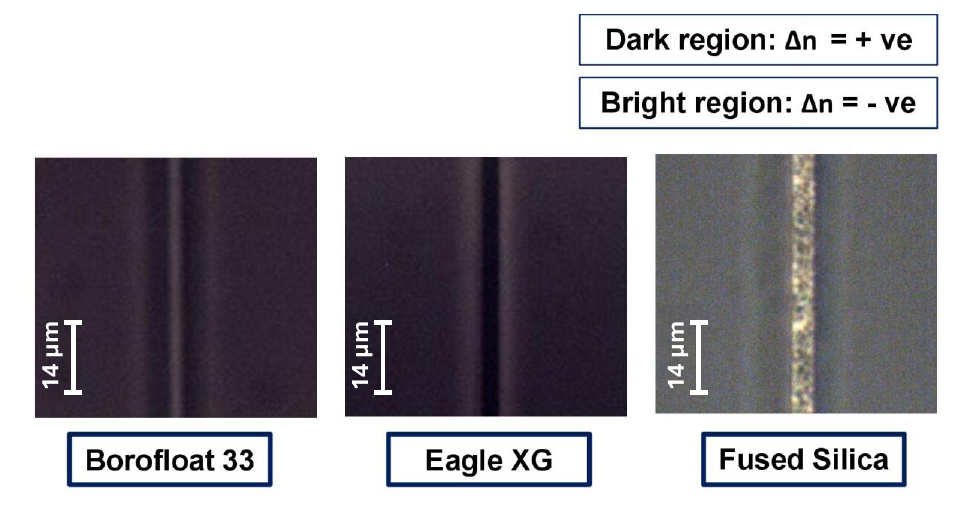}
    \caption{PCM images of ULI in different glass materials. Borofloat 33 and Fused silica exhibit Type II modifications, while Eagle XG shows a Type I modification.}
    \label{fig:materials response summary}
\end{figure}

\noindent \textcolor{black}{As an initial screening step, the laser-induced modifications in each material were examined using phase-contrast microscopy (PCM). Borofloat 33 exhibited a negative refractive-index modification (Type II), characterized by a reduced on-axis refractive index relative to the surrounding material. Fused silica similarly displayed a Type II response. However, the modified region was accompanied by spatially non-uniform, crack-like features indicative of localized material damage. In contrast, Eagle XG demonstrated a positive refractive-index modification (Type I), characterized by an increase in the on-axis refractive index and a more homogeneous profile of modification, as shown in Figure~\ref{fig:materials response summary}. Based on these observations, Eagle XG was identified as the most promising candidate for subsequent optimization of the ULI parameters and detailed optical characterization.}

\begin{figure} [ht!]
     \centering
     \includegraphics[width=0.7\linewidth]{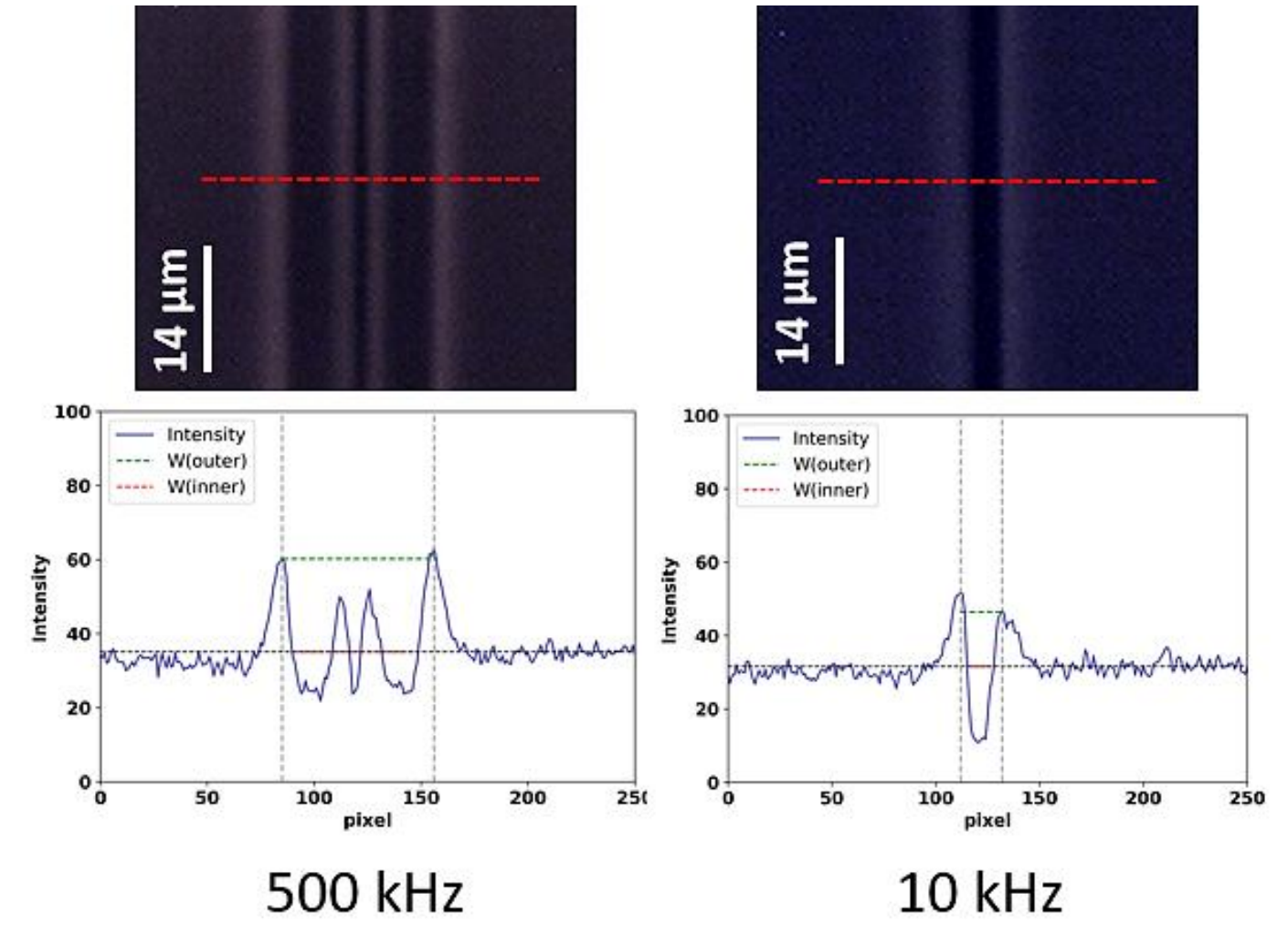}
     \caption{Index profile analysis of PCM images obtained by extracting the pixel-intensity distribution across the cross-section of the modified region for waveguides fabricated at high and low repetition rates.}
     \label{fig:analysis}
 \end{figure}

\noindent \textcolor{black}{Given its positive refractive-index response and comparatively uniform modification profile, Eagle XG was selected for further optimization and detailed characterization. To obtain a preliminary quantitative assessment of the laser-induced modifications, phase contrast microscopy (PCM) images of the inscribed structures were analyzed. Intensity profiles were extracted along selected cross-sections of the modified regions, and the resulting pixel-intensity distributions were used to estimate the modification widths, as illustrated in Figure~\ref{fig:analysis}. This approach provided a rapid means of comparing the effects of different inscription parameters and identifying trends in the waveguide.}

% \textcolor{black}{Given its positive refractive-index response and comparatively uniform modification profile, Eagle XG was selected for further optimization and detailed characterization. To obtain a preliminary quantitative assessment of the laser-induced modifications, phase-contrast microscopy (PCM) images of the inscribed structures were analyzed. Intensity profiles were extracted along selected cross-sections of the modified regions, and the resulting pixel-intensity distributions were used to estimate the characteristic modification widths, as illustrated in Figure~\ref{fig:analysis}. This approach provided a rapid means of comparing the effects of different inscription parameters and identifying trends in the waveguide morphology.}

 \begin{figure} [ht!]
    \centering
    \includegraphics[width=0.55\linewidth]{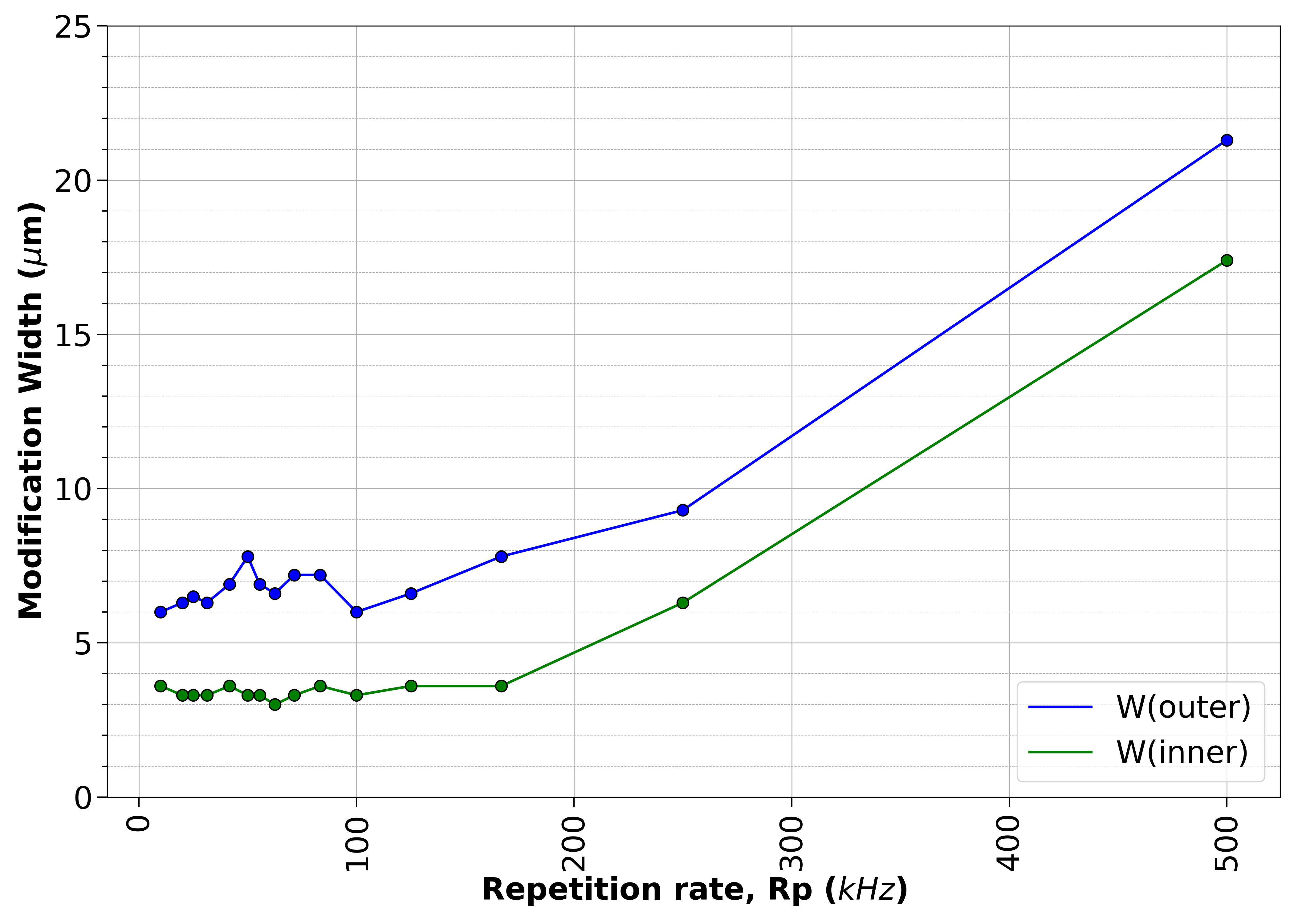}
    \caption{Cross-sectional size of the laser-induced modification as a function of repetition rate for waveguides fabricated in Eagle XG. The modification width increases with increasing repetition rate, indicating a larger modified region at higher repetition rates.}
    \label{fig:rp test}
\end{figure}

\noindent \textcolor{black}{It should be noted that the analysis was performed using top-view PCM images. As a result, the measured widths do not necessarily correspond to the true cross-sectional dimensions of the waveguides. Nevertheless, the extracted values provide a useful first-order metric for comparing the relative extent of the modified regions and assessing their dependence on the fabrication conditions.} \textcolor{black}{For consistency throughout this work, the outer width is defined as the distance between the two intensity maxima, corresponding to the full extent of the laser-modified region. The inner width is defined as the distance between the locations where the two intensity maxima intersect the central intensity minimum at the background intensity level.}

\noindent \textcolor{black}{The analysis revealed a clear dependence of the modification dimensions on the inscription parameters. In particular, reducing the laser repetition rate resulted in narrower modification widths and a noticeable reduction in crack-like features within the modified region. Using the calculation procedure shown in Figure~\ref{fig:analysis}, the effect of modification dimensions with respect to laser repetition rate is shown in Figure~\ref{fig:rp test}. This behavior suggests that the lower repetition rate reduces cumulative thermal effects during inscription, leading to improved structural homogeneity in the waveguides. Consequently, repetition rate emerged as a key parameter in the optimization process and was investigated further in subsequent experiments.}

\noindent \textcolor{black}{While the PCM analysis provided a rapid indication of the quality of the laser-induced modifications, confirmation of waveguide performance ultimately requires optical characterization. Therefore, the fabricated structures were evaluated at near-infrared wavelengths using the characterization bench described in Figure~\ref{fig:IR-bench}. A 1310 nm laser source was coupled into the waveguides, and efficient light confinement and propagation were observed in structures fabricated within Eagle XG glass. These results are consistent with the positive refractive-index modifications identified in the PCM.}

\noindent \textcolor{black}{A systematic comparison of the output mode profiles revealed that the majority of the fabricated waveguides supported multimode propagation at 1310 nm. However, waveguides inscribed at repetition rates of 50 kHz and below consistently exhibited near-single-mode behavior. These structures generally produced more circular output modes, although their mode field diameters (MFDs) were larger than those of the reference SMF-28 fiber. In contrast, waveguides fabricated at 25 kHz displayed slightly reduced mode circularity but achieved mode sizes that more closely matched the nominal MFD of SMF-28, an important consideration for efficient fiber-to-chip coupling.}

\begin{figure} [ht!]
    \centering
    \includegraphics[width=0.65\linewidth]{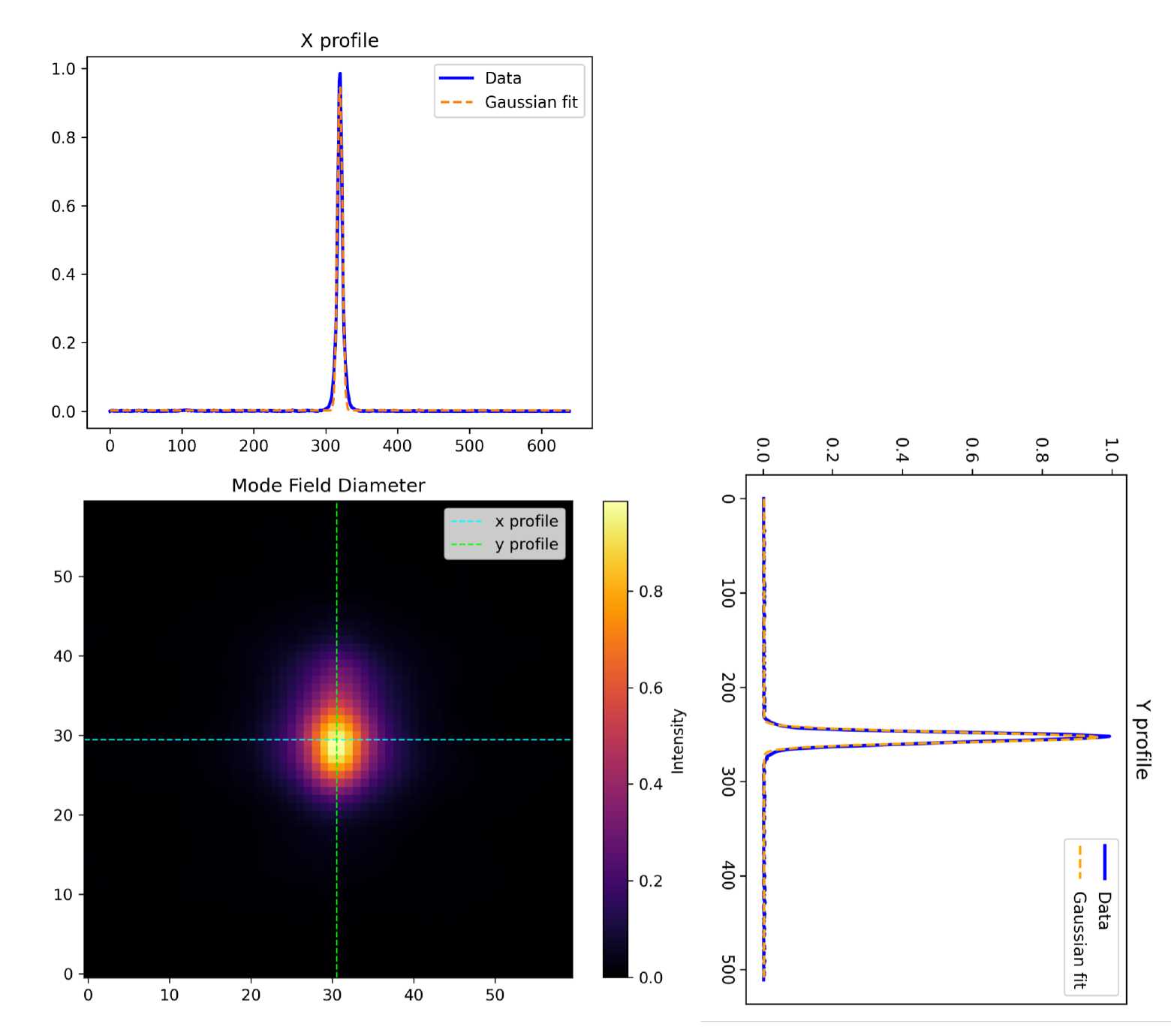}
    \caption{Mode field diameter of the fabricated waveguide at a repetition rate of 25 kHz, showing the measured intensity distribution and corresponding Gaussian fits along the horizontal (x) and vertical (y) axes. This structure exhibits the best single-mode behavior among the investigated waveguides, with mode field dimensions of 7.67~$\mu$m $\times$ 11.41~$\mu$m.}
    \label{fig:MFD}
\end{figure}

\noindent \textcolor{black}{The best-performing waveguide identified in this study was fabricated at a repetition rate of 25 kHz. From Figure~\ref{fig:MFD}, the measured mode dimensions were determined to be 7.67~\textmu m $\times$ 11.41~\textmu m, corresponding to a mode circularity of 67\%. The measured mode dimensions differed by less than 24\% from the nominal MFD of the SMF-28 fiber\cite{corning_smf28}. The inscription conditions associated with the most favorable waveguide performance were a pulse energy of approximately 475 nJ, and scan speeds 2 mms$^{-1}$, corresponding to a representative net fluence of approximately 46.12~$\mu$J $\mu$m$^{-2}$.}

% The best-performing waveguide identified in this study was fabricated at a repetition rate of 25 kHz and exhibited mode field dimensions of 7.67~\textmu m $\times$ 11.41~\textmu m, corresponding to a mode circularity of 67\%. As shown in Figure~\ref{fig:MFD}, the measured mode dimensions differed by less than 24\% from the nominal MFD of the SMF-28 fiber.
% From Figure 8, the measured mode dimensions...
% coupling lgihjt fingsings

\noindent These results indicate that lower repetition-rate inscription conditions are beneficial for achieving waveguides with mode characteristics suitable for integration with standard single-mode optical fibers in future J-band astrophotonic devices.

% This structure exhibits the best single-mode behavior among the investigated waveguides.

% % checking index using other method to find neumerical value and check the validity 

% Further optimization of waveguides fabricated at repetition rates below 25~kHz could potentially lead to improved single-mode operation with reduced MFDs more compatible with \textit{SMF-28} fibre. Additionally, beam profiling of the fs-laser beam could improve the circularity of the waveguide cross-section and enhance modal symmetry.
% requires more work, 
% cite lucas paper
\newpage
%------------------
\section{Conclusion and Future Work}
\label{sec: future work}  % \label{} allows reference to this section

\textcolor{black}{This work represents the first step toward establishing an in-house ultrafast laser inscription (ULI) capability at AIP for the development of astrophotonic components operating in the astronomical J-band. A systematic workflow for fabrication and characterization was implemented, combining rapid phase-contrast microscopy (PCM) screening with near-infrared optical characterization to identify suitable substrate materials and inscription parameters for waveguide fabrication.}

\noindent \textcolor{black}{PCM proved to be an effective tool for the rapid assessment of laser-induced modifications, enabling efficient comparison of different materials and fabrication conditions. Among the three investigated substrates, Eagle XG exhibited a positive refractive-index modification, making it the most promising candidate for optical waveguide fabrication. The presence of this positive index change was subsequently confirmed by successfully guiding 1310 nm light, demonstrating the ability of the laser-written structures to support near-infrared propagation.}

\noindent \textcolor{black}{Optical characterization revealed that the majority of the fabricated waveguides exhibited multimode behavior, particularly at repetition rates above 50 kHz. However, near-single-mode operation was consistently observed for waveguides fabricated at the lower repetition rate. The most favorable performance was achieved at a repetition rate of 25 kHz, where the resulting mode field dimensions of 7.67~\textmu m $\times$ 11.41~\textmu m differed by less than 24\% from the nominal mode field diameter of the SMF-28 reference fiber. These results indicate that low-repetition-rate inscription conditions, combined with a pulse energy of 475 nJ and scan speed 2 mm s$^{-1}$, provide a promising route toward the fabrication of single-mode waveguides suitable for integration with existing astrophotonic components.}

\noindent \textcolor{black}{Building on these initial results, future work will focus on validating the reproducibility and long-term stability of the identified fabrication parameters. Additional characterization techniques will be employed to directly quantify the induced refractive-index change and to better understand the relationship between inscription conditions and waveguide performance. Furthermore, advanced fabrication strategies, including slit beam shaping and multi-scan inscription approaches, will be investigated to improve mode symmetry and achieve waveguides with more circular cross-sections and enhanced coupling efficiency to standard single-mode fibers.}

\noindent \textcolor{black}{A more comprehensive optical characterization will also be undertaken, including measurements of throughput, propagation loss, and coupling efficiency. Ultimately, the optimized fabrication process developed in this work will serve as the foundation for realizing more complex ULI-fabricated astrophotonic devices at AIP, including integrated-optics beam combiners and other components for future astronomical interferometry and high-angular-resolution instrumentation.}

\acknowledgments % equivalent to \section*{ACKNOWLEDGMENTS}       
\textcolor{black}{R.I. sincerely thanks SPIE for providing a Student Support Grant, which enabled participation in and presentation of this work at the SPIE Astronomical Telescopes + Instrumentation conference 2026. The valuable discussions, guidance, and technical insights provided by Dr. Ahmed Sanny and Dr. Alexandre Mermillod-Blondin during various stages of this project are also gratefully acknowledged. The authors gratefully acknowledge financial support from the Deutsche Forschungsgemeinschaft (DFG) through the APPEXIS project (grant no. 506421303) at the Leibniz Institute for Astrophysics Potsdam (AIP).}

% References
\bibliography{report} % bibliography data in report.bib
\bibliographystyle{spiebib} % makes bibtex use spiebib.bst

\end{document}